\documentclass[12pt]{article}
\usepackage{amsmath,amsfonts,latexsym}

\def\01{\{0,1\}}
\newcommand{\ceil}[1]{\lceil{#1}\rceil}
\newcommand{\ket}[1]{|#1\rangle}

\newcommand{\eps}{\varepsilon} 
\newcommand{\Nr}{\left\lceil\frac{N}{r}\right\rceil}
\newcommand{\set}[1]{\left[#1\right]}
\newcommand{\abs}[1]{\left|#1\right|}

\newcommand{\ED}{\textup{\textbf{\textrm{ED}}}}
\newcommand{\Claw}{\textup{\textbf{\textrm{Claw}}}}
\newcommand{\Triangle}{\textup{\textbf{\textrm{Triangle}}}}
\newcommand{\OR}{\textup{\textbf{\textrm{OR}}}}

\newtheorem{theorem}{Theorem}
\newtheorem{lemma}[theorem]{Lemma}
\newtheorem{definition}[theorem]{Definition}

\newcommand{\smallspace}{\mskip 2mu minus 1mu}

\newcommand{\tinyspace}{\mskip 1mu}

\newenvironment{proof}
{\noindent {\bf Proof }}
{{\hfill $\Box$}\\
\smallskip
}

\newenvironment{algorithm}[1]{\medskip\noindent%
\itemsep0pt\begin{trivlist}\item[]%
{\flushleft\textbf{Algorithm #1}}
\begin{enumerate}}%
{\end{enumerate}\end{trivlist}\medskip}

\begin{document}

\title{Quantum Algorithms for Element Distinctness%
\thanks{\,Research partially supported by 
the EU fifth framework program QAIP, IST-1999-11234.
\smallskip}}

\author{%
\makebox[3.6cm]{Harry Buhrman%
\thanks{\,CWI, P.O.~Box~94079, Amsterdam, The Netherlands.
 email:~\mbox{\texttt{buhrman}\textbf{\char"40}\texttt{cwi.nl}}.}}
\and 
\addtocounter{footnote}{1}
\makebox[3.6cm]{Christoph D{\"u}rr$\smallspace$%
\thanks{\,Universit{\'e} Paris-Sud, LRI, 91405 Orsay, France.
email:~\mbox{\texttt{durr}\textbf{\char"40}\texttt{lri.fr}}.}}
\and 
\makebox[3.6cm]{Mark Heiligman$\smallspace$%
\thanks{\,NSA, Suite 6111, Fort George G.~Meade, \mbox{MD~20755.
email:~{\texttt{mheilig}\textbf{\char"40}\texttt{zombie.ncsc.mil}}.}}}
\and 
\makebox[3.6cm]{Peter H{\o}yer$\smallspace$%
\thanks{\,BRICS, University of Aarhus, Aarhus~C, Denmark.
email:~\mbox{\texttt{hoyer}\textbf{\char"40}\texttt{brics.dk}}.}}
\and 
\makebox[3.6cm]{Fr{\'e}d{\'e}ric Magniez$\smallspace$%
\thanks{\,CNRS, Universit\'e Paris-Sud, LRI, 91405 Orsay, France.
email:~\mbox{\texttt{magniez}\textbf{\char"40}\texttt{lri.fr}}.}}
\and 
\makebox[3.6cm]{Miklos Santha$\smallspace$%
\thanks{\,CNRS, Universit{\'e} Paris-Sud, LRI, 91405 Orsay, France.
email:~\mbox{\texttt{santha}\textbf{\char"40}\texttt{lri.fr}}.}
}\\[.001cm]
\and 
\makebox[3.6cm]{Ronald de~Wolf$\smallspace$%
\thanks{\,CWI, P.O.~Box~94079, Amsterdam, The Netherlands.
email:~\mbox{\texttt{rdewolf}\textbf{\char"40}\texttt{cwi.nl}}.}}}

\date{September~1, 2000}

\setcounter{footnote}{-1}
\maketitle


\begin{abstract}
We~present several applications of quantum amplitude amplification 
to finding claws and collisions in ordered or unordered functions. 
Our algorithms generalize those of Brassard, H{\o}yer, and Tapp,
and imply an $O(N^{3/4}\log N)$ quantum upper bound for the element 
distinctness problem in the comparison complexity model
(contrasting with $\Theta(N\log N)$ classical complexity).
We also prove a lower bound of $\Omega(\sqrt{N})$ comparisons 
for this problem and derive bounds for a number of related problems.
\end{abstract}

\section{Introduction}
In the last decade, quantum computing has become a
prominent and promising area of theoretical computer science.
Realizing this promise requires two things:
(1)~actually building a quantum computer and (2)~discovering tasks 
where a quantum computer is significantly faster 
than a classical computer.
Here we are concerned with the second issue.
Few good quantum algorithms are known to date, the two main examples 
being Shor's algorithm for factoring~\cite{shor:factoring} 
and Grover's algorithm for searching~\cite{grover:search}.
Whereas the first so far has remained a somewhat isolated
although seminal result, the second 
has been applied as a building block in quite a few other quantum 
algo\-rithms~\cite{brassard&hoyer:simon,bht:collision,BuhrmanCleveWigderson98,bcwz:qerror,nayak&wu:median,bhmt:countingj}.

One of the earliest applications of Grover's algorithm was 
the algorithm of
Brassard, H{\o}yer, and Tapp~\cite{bht:collision} for finding 
a \emph{collision} in a 2-to-1 function~$f$. 
A collision is a pair of distinct elements $x,y$ 
such that \mbox{$f(x)=f(y)$}. 
Suppose the size of $f$'s domain is~$N$. 
For a classical randomized algorithm, $\Theta(\sqrt{N})$
evaluations of the function are necessary and sufficient 
to find a collision.
The quantum algorithm of~\cite{bht:collision}
finds a collision using $O(N^{1/3})$ evaluations of~$f$.
No non trivial quantum lower bound is known for this problem.
A notion related to collisions is that of a \emph{claw}.
A claw in functions $f$ and $g$ is a pair $(x,y)$ such that $f(x)=g(y)$.
If $f$ and $g$ are permutations on $[N]=\{1,\ldots,N\}$, then 
the function on $[2N]$ which maps the first half of the domain
according to $f$ and the second half according to $g$, is a 2-to-1 function.
Thus the algorithm of Brassard, H{\o}yer, and Tapp can also find a claw 
in such $f$ and $g$ using $O(N^{1/3})$ evaluations of $f$ and~$g$.  

In this paper we consider the quantum complexity of collision-finding
or claw-finding with and without restrictions on the 
functions $f$ and~$g$.
In Section~\ref{secbothunordered} we consider the situation where
$f:[N]\rightarrow Z$ and $g:[M]\rightarrow Z$ are arbitrary.
Our aim is to find a claw between $f$ and $g$, if one exists.
For now, let us assume $N=M$ (in the body of the paper we treat the general case).
The complexity measure we use is the number of \emph{comparisons} between elements.
That is, we assume a total order on $Z$
and our only way to access $f$ and $g$ is by comparing $f(x)$ with $f(y)$,
$g(x)$ with $g(y)$, or $f(x)$ with $g(y)$, according to this total order.
The ability to make such comparisons is weaker than the ability 
to evaluate and actually learn the function values $f(x)$ and $g(y)$.
However, our bounds remain the same up to logarithmic factors if we 
were to count the number of
\emph{function-evaluations} instead of comparisons.

An optimal \emph{classical} algorithm for this general claw-finding problem 
is the following. Viewing $f$ as a list of $N$ items, we can sort it using 
$N\log N+O(N)$ comparisons.  Once $f$ is sorted, we can for a given 
$y\in[N]$ find an $x$ such that $f(x)=g(y)$ 
provided such an $x$ exists,
using $\log N$ comparisons (by utilizing binary search on~$f$).
Thus exhaustive search on all $y$ yields
an $O(N\log N)$ algorithm for finding a claw with certainty,
provided one exists.
This $N\log N$ is optimal up to constant factors even for bounded-error
classical algorithms. 
Here we show that a quantum computer can do better:
we exhibit a quantum algorithm that finds a claw 
with high probability using $O(N^{3/4}\log N)$ comparisons.
We also prove a lower bound for this problem of $\Omega(N^{1/2})$ 
comparisons for bounded-error
quantum algorithms and $\Omega(N)$ for exact quantum algorithms.

Our algorithm for claw-finding also yields an $O(N^{3/4}\log N)$ 
bounded-error quantum algorithm for finding a collision for 
arbitrary functions.
Note that \emph{deciding} if a collision occurs in $f$ 
is equivalent to deciding whether
$f$ maps all $x$ to distinct elements. This is known as the 
\emph{element distinctness} problem and has been well studied classically, 
see e.g.~\cite{yao:timespace,lubiw&racz:distinctness,grigoriev:lower,bssv:timespace}.
Element distinctness is particularly interesting because its classical 
complexity is related to that of sorting, which is well known to require 
$N\log N+\Theta(N)$ comparisons. 
If we sort $f$, we can decide element
distinctness by going through the sorted list once,
which gives a classical upper bound of $N\log N+O(N)$ comparisons.
Conversely, element distinctness requires $\Omega(N\log N)$ comparisons
in case of classical bounded-error algorithms 
(even in a much stronger model, see~\cite{grigoriev:lower}),
so sorting and element distinctness are equally hard 
for classical computers.
On a quantum computer, the best known upper bound 
for sorting is roughly $0.53\ N\log N$ 
compa\-ri\-sons~\cite{fggs:binsearchupper}, whereas
the best known lower bound is 
$\Omega(N)$~\cite{fggs:distinguish}.
Accordingly, our $O(N^{3/4}\log N)$ quantum upper bound 
shows that element distinctness is significantly easier than sorting 
for a quantum computer, in contrast to the classical case.

In Section~\ref{secfordered}, we consider the case where $f$ is ordered
(monotone non-decreasing): $f(1)\leq f(2)\leq\cdots\leq f(N)$.
In this case, the quantum complexity of claw-finding and collision finding
drops from $O(N^{3/4}\log N)$ to $O(N^{1/2}\log N)$.
In Section~\ref{secfgordered} we show how to remove the $\log N$ factor
(replacing it by a near-constant function) 
if both $f$ and $g$ are ordered.
The lower bound for this restricted case remains $\Omega({N}^{1/2})$.

In Section~\ref{sechardproblems} we give some problems
related to the element distinctness problem for 
which quantum computers cannot help.
We then, in Section~\ref{sectrianglefinding}, 
give bounds for the number of edges a quantum computer 
needs to query in order to find a \emph{triangle}
in a given graph (which, informally, can be viewed as 
a collision between three nodes).
Finally, we end with some concluding remarks in 
Section~\ref{secohmydearthemoonisagreencheese}.

\section{Preliminaries}

We consider the following problems:

\begin{description}
\item[\textbf{Claw-finding problem}]\mbox{}\\
Given two functions $f: X\rightarrow Z$ and $g: Y\rightarrow Z$,
find a pair $(x,y)\in X\times Y$ such that $f(x)=g(y)$.
\item[\textbf{Collision-finding problem}]\mbox{}\\
Given a function $f: X\rightarrow Z$, 
find two distinct elements $x,y \in X$ such that $f(x)=f(y)$.
\end{description}

We assume that $X=[N]=\{1,\ldots, N\}$ and 
$Y=[M]=\{1,\ldots,M\}$ with $N\leq M$.

For details about quantum computing we refer 
to~\cite{berthiaume:qc,cleve:intro}.
We formalize a comparison between $f(x)$ and $f(y)$ as an application 
of the following unitary transformation:
$$
\ket{x,y,b}\;\longmapsto\;\ket{x,y,b\oplus[f(x)\leq f(y)]},
$$
where $b\in\01$ and $[f(x)\leq f(y)]$ 
denotes the truth-value of the statement ``$f(x)\leq f(y)$''.
We~formalize comparisons between $f(x)$ and $g(y)$ similarly.

We are interested in the number of comparisons required for
claw-finding or collision-finding. We use $Q_E(P)$ and $Q_2(P)$
for the worst-case number of comparisons required for solving problem
$P$ by exact and bounded-error quantum algorithms, respectively.
In our algorithms we make abundant use of quantum 
\emph{amplitude amplification}~\cite{bhmt:countingj},
which generalizes quantum search~\cite{grover:search}.
The essence of amplitude amplification can be summarized
by following theorem.
\begin{theorem}[Amplitude amplification]
There exists a quantum algorithm {\bf QSearch} with the following property.
Let $\cal A$ be any quantum algorithm that uses no measurements, and let
$\chi:\mathbb{Z}\rightarrow\01$ be any Boolean function.
Let $a$ denote the initial success probability of $\cal A$
of finding a solution
(i.e.~the probability of outputting $z$ s.t.~$\chi(z)=1$).
Algorithm {\bf QSearch} finds a solution using an expected number of
$O(1/\sqrt{a})$ applications of $\cal A$ and ${\cal A}^{-1}$ if $a>0$,
and otherwise runs forever.
\end{theorem}

The algorithm {\bf QSearch} does not need to know the value of $a$
in advance, but if $a$ is known, it can find a solution in
\emph{worst-case} $O(1/\sqrt{a})$ applications.

Grover's algorithm for searching a space of $N$ items is a special 
case of amplitude amplification. 
If $\cal A$ is the classical algorithm which
selects a random element of the space and checks it, then $a=1/N$, and
amplitude amplification implies an $O({N}^{1/2})$ quantum algorithm 
for searching the space.  
We~refer to this process as ``quantum searching''.

\section{Finding claws if $f$ and $g$ are not ordered}\label{secbothunordered}

First we consider the most general case, where $f$ and $g$
are arbitrary, possibly unordered functions.
Our claw-finding algorithms are instances of the 
following generic algorithm, which is parameterized by an integer
$\ell\leq\min\{N,\sqrt{M}\}$:
\begin{algorithm}{Generic claw-finder}
\item Select a random subset $A\subseteq[N]$ of size $\ell$ 
      \label{generic:A}
\item Select a random subset $B\subseteq[M]$ of size $\ell^2$
      \label{generic:B}
\item Sort the elements in $A$ according to their $f$-value
      \label{generic:sort}
\item 
For a specific $b\in B$, we can check if there is an $a\in A$ 
such that $(a,b)$ is a claw using classical binary search on 
the sorted version of $A$.
Combine this with quantum search on the $B$-elements to search for a claw
in $A\times B$.
      \label{generic:search}
\item Apply amplitude amplification on steps 
      \mbox{\ref{generic:A}--\ref{generic:search}}
      \label{generic:amplify}
\end{algorithm}
We analyze the comparison-complexity of this algorithm.
Step~\ref{generic:sort} 
just employs classical sorting and hence takes $\ell\log\ell+O(\ell)$
comparisons. Step~\ref{generic:search} takes $O(\sqrt{|B|}\log|A|)=O(\ell\log\ell)$ 
comparisons, so steps \mbox{\ref{generic:A}--\ref{generic:search}}
take $O(\ell\log\ell)$ comparisons in total.

If no claws between $f$ and $g$ exist, then this algorithm
does not terminate. 
Now suppose there is a claw $(x,y)\in X\times Y$.
Then $(x,y)\in A\times B$ with probability $(\ell/N)\cdot(\ell^2/M)$, 
and if indeed $(x,y)\in A\times B$, then step~\ref{generic:search} 
will find this (or some other) collision 
with probability at least~$1/2$ in at most $O(\ell \log \ell)$
comparisons.
Hence the overall success probability of 
steps \mbox{\ref{generic:A}--\ref{generic:search}} is at least 
$a=\ell^3/2NM$, and the amplitude amplification
of step~5 requires an expected number of $O(\sqrt{NM/\ell^3})$ 
iterations of steps \mbox{\ref{generic:A}--\ref{generic:search}}.
Accordingly, the total expected number of comparisons to find
a claw is $O(\sqrt{\frac{NM}{\ell}}\log\ell)$,
provided there is one.
In order to minimize the number of comparisons, we maximize $\ell$, 
subject to the constraint $\ell\leq\min\{N,\sqrt{M}\}$.
This gives upper bounds of $O(N^{1/2}M^{1/4}\log N)$ comparisons
if $N\leq M\leq N^2$, and $O(M^{1/2}\log N)$ if $M>N^2$.

What about \emph{lower} bounds for the claw-finding problem?
We can reduce the $\OR$-problem to claw-finding as follows.
Given a function $g:[M]\rightarrow\01$,
by definition $\OR(g)$ is 1 if there is an $i$ such that $g(i)=1$.
To determine this value, we set $N=1$ and
define $f(1)=1$.
Then there is a claw between $f$ and $g$
if and only if $\OR(g)=1$.
Thus if we can find a claw using $c$ comparisons, we can decide $\OR$
using $2c$ queries to $g$ (two $g$-queries suffice to implement a comparison).
Using known lower bounds for the $\OR$-function~\cite{bbbv:str&weak,bbcmw:polynomials}, 
this gives an 
$\Omega(M)$ bound for exact quantum and  an $\Omega(\sqrt{M})$ bound  for 
bounded-error quantum algorithms. 
The next theorem follows.
\begin{theorem}
The comparison-complexity of the claw-finding problem is
\begin{itemize}
\item $\Omega(M^{1/2})\leq Q_2(\Claw)\leq \displaystyle
\left\{ \begin{array}{ll}
            O(N^{1/2}M^{1/4}\log N) & \mbox{if $N\leq M\leq N^2$}\\
            O(M^{1/2}\log N)        & \mbox{if $M>N^2$}
        \end{array}\right.$
\item $\Omega(M)\leq Q_E(\Claw)\leq O(M\log N)$.
\end{itemize}
\end{theorem}

The bounds for the case $M>N^2$ and the case of exact 
computation are tight up to the $\log N$ term, but 
the case $M\leq N^2$ is nowhere near tight.
In particular, for $N=M$ the complexity lies somewhere 
between $N^{1/2}$ and $N^{3/4}\log N$.

Now consider the problem of finding a collision for an 
arbitrary function $f:[N]\rightarrow Z$.
A simple modification of the above algorithm for claw-finding
works fine to find such $(x,y)$-pairs if they exist
(put $g=f$ and avoid claws of the form $(x,x)$), 
and gives that $Q_2(\ED)\in O(N^{3/4}\log N)$.
The best known lower bounds follow again via reductions 
from the $\OR$-problem:
given $X\in\01^N$, we define $f:[N+1]\rightarrow\{0,\ldots,N\}$ 
as $f(i)=i(1-x_i)$ and $f(N+1)=0$. Now $\OR(X)=1$ if
and only if $f$ contains a collision.
As mentioned in the introduction, the problem of deciding if there
is a collision is equivalent to the element distinctness (ED) problem.
\pagebreak[4]
Theorem~\ref{thm:ed} follows.
\begin{theorem}\label{thm:ed}
The comparison-complexity of the element distinctness problem is
\begin{itemize}
\item $\Omega(N^{1/2})\leq Q_2(\ED)\leq O(N^{3/4}\log N)$
\item $\Omega(N)\leq Q_E(\ED)\leq O(N\log N)$.
\end{itemize}
\end{theorem}

In contrast, for classical (exact or bounded-error) algorithms,
element distinctness is as hard as sorting and requires
$\Theta(N\log N)$ comparisons.

Collision-finding becomes cheaper if we know that some value $z\in Z$ 
occurs at least $k$ times. 
If we pick a random subset $S$ of $10N/k$ of the domain,
then with high probability at least two pre-images of $z$ will
be contained in $S$. Thus running our algorithm on $S$ will find
a collision with high probability, resulting in complexity 
$O((N/k)^{3/4}\log(N/k))$.
Also, if $f$ is a 2-to-1 function, we can rederive the $O(N^{1/3}\log N)$
bound of Brassard, H{\o}yer, and Tapp~\cite{bht:collision} 
by taking $\ell=N^{1/3}$. This yields constant success probability 
after steps \mbox{\ref{generic:A}--\ref{generic:search}}
in the generic algorithm, and hence no further rounds of amplitude 
amplification are required.
As in the case of~\cite{bht:collision}, this algorithm 
can be made exact by using the exact form of amplitude amplification 
(the success probability can be exactly computed in this case).

\section{Finding claws if $f$ is ordered}\label{secfordered}

Now suppose that function $f$ is 
ordered: $f(1)\leq f(2)\leq\cdots\leq f(N)$,
and that function $g : [M] \rightarrow Z$ is \emph{not} necessarily
ordered.
In this case, given some $y\in[M]$, we can find an $x \in [N]$ 
such that
$(x,y)$ is a claw using binary search on~$f$.
Thus, combining this with a quantum search on all $y \in [M]$, 
we obtain the upper bound of $O(\sqrt{M}\log N)$ for finding 
a claw in $f$ and~$g$.
The lower bounds of the last section via the $\OR$-reduction 
still apply, and hence we obtain the following theorem.
\begin{theorem}
The comparison-complexity of the claw-finding problem with ordered $f$ is
\begin{itemize}
\item $\Omega(M^{1/2})\leq Q_2(\Claw)\leq O(M^{1/2}\log N)$ 
\item $\Omega(M)\leq Q_E(\Claw)\leq O(M\log N)$.
\end{itemize}
\end{theorem}

Note that collision-finding for an ordered $f:[N]\rightarrow Z$
is equivalent to searching a space of $N-1$ items (namely all 
consecutive pairs in the domain of $f$) 
and hence requires $\Theta(\sqrt{N})$ comparisons.

\section{Finding claws if both $f$ and $g$ are ordered}\label{secfgordered}

Now consider the case where both $f$ and $g$ are ordered.
Assume for simplicity that $N=M$.
Again we get an $\Omega(\sqrt{N})$ lower bound via a reduction 
from the $\OR$-problem as follows.
Given an $\OR$-instance $X\in\01^N$, we define $f,g: \set{N} \rightarrow Z$ 
by $f(i) = 2i+1$ and $g(i) = 2i + x_i$ for all $i \in \set{N}$.
Then $f$ and $g$ are ordered, and $\OR(X)=1$ if and only if 
there is a claw between $f$
and $g$. The lower bound follows.

We give a quantum algorithm that solves the problem using 
$O\big(\sqrt{N} \tinyspace c^{\log^\star(N)}\big)$ comparisons
for some constant $c > 0 $.
The function $\log^\star(N)$ is defined as the minimum number of iterated applications
of the logarithm function necessary to obtain a number less than or equal to~1:
$\log^\star(N) = \min\{i \geq 0\mid \log^{(i)}(N) \leq 1\}$, 
where $\log^{(i)} = \log \circ \log^{(i-1)}$ denotes the $i$th 
iterated application of $\log$, and $\log^{(0)}$ is the identity function.
Even though $c^{\log^\star(N)}$ is exponential
in~$\log^\star(N)$, it is still very small in~$N$,
in particular $c^{\log^\star(N)} \in o(\log^{(i)}(N))$ for any constant~$i\geq 1$.
Thus we replace the $\log N$ in the upper bound of the previous section
by a near-constant function.
Our algorithm defines a set of 
subproblems such that the original problem $(f,g)$ contains
a claw if and only if at least one of the subproblems contains a claw.  
We then solve the original problem by running the subproblems 
in quantum parallel and applying amplitude amplification.

Let $r>0$ be an integer.  We~define $2 \Nr$ subproblems as follows.

\begin{definition}
Let $r>0$ be an integer and $f,g : \set{N} \rightarrow Z$.

For each $0\leq i\leq \ceil{N/r}-1$,
we define the subproblem $(f_i,g'_i)$ by letting 
$f_i$ denote the restriction of $f$ to subdomain $[ir+1,(i+1)r]$, 
and $g'_i$ the restriction of $g$ to~$[j,j+r-1]$ where 
$j$ is the minimum $j' \in \set{N}$ such that $g(j') \geq f(ir+1)$.

Similarly, for each $0\leq j\leq \ceil{N/r}-1$,
we define the subproblem $(f'_j,g_j)$ by letting 
$g_j$ denote the restriction of $g$ to~$[jr+1,(j+1)r]$, and
$f'_j$ the restriction of $f$ to~$[i,i+r-1]$ where 
$i$ is the minimum $i' \in \set{N}$ such that $f(i') \geq g(jr+1)$.
\end{definition}

It is not hard to check that these subproblems all together
provide a solution to the original problem.

\begin{lemma}
Let $r>0$ be an integer and $f,g : \set{N} \rightarrow Z$.
Then $(f,g)$ contains a claw if and only if
for some $i$ or $j$ in $[0,\ceil{N/r}-1]$
the subproblem $(f_i,g'_i)$ or $(f'_j,g_j)$ contains a claw.
\end{lemma}

Each of these $2 \Nr$ subproblems 
is itself an instance of the claw-finding problem of size~$r$.
By running them all together in quantum parallel and
then applying amplitude amplification, we obtain our main result.
\begin{theorem}\label{thm:mon}
There exists a quantum algorithm that outputs a claw between $f$ and $g$ 
with probability at least $\frac23$ provided one exists, using 
$O\big(\sqrt{N} \tinyspace c^{\log^\star(N)}\big)$ comparisons, for some constant~$c$.
\end{theorem}

\begin{proof}
Let $T(N)$ denote the worst-case number of comparisons required
if $f$ and~$g$ have domain of size $N$. We show that
\begin{equation}\label{eq:cost}
T(N) \smallspace\leq\smallspace 
c' \sqrt{\frac{N}{r}}\smallspace \bigg(\lceil\log(N+1)\rceil+T(r)\bigg),
\end{equation}
for some (small) constant $c'$.
Let $0\leq i\leq \ceil{N/r}-1$ and consider the subproblem $(f_i,g'_i)$.
Using at most $\lceil\log(N+1)\rceil+T(r)$ comparisons,
we can find a claw in $(f_i,g'_i)$ with probability at
least~$\frac{2}{3}$, provided there is one.  
We~do that by using binary search to find the minimum~$j$ 
for which $g(j) \geq f(ir+1)$, at the cost of $\lceil\log(N+1)\rceil$ 
comparisons, and then recursively determining if the 
subproblem $(f_i,g'_i)$ contains a claw at the cost of at 
most $T(r)$ additional comparisons.
There are $2 \Nr$ subproblems, so by~applying amplitude amplification we 
can find a claw among any one of them with probability at least~$\frac{2}{3}$, 
provided there is one, in the number of comparisons given in equation~(\ref{eq:cost}).

We~pick $r = \lceil \log^2(N)\rceil$.
Since $T(r)\geq \Omega(\sqrt{r})=\Omega(\log N)$, 
equation~(\ref{eq:cost}) implies
\begin{equation}\label{eq:cost'}
T(N) \smallspace\leq\smallspace c'' \sqrt{\frac{N}{r}}T(r),
\end{equation}
for some constant $c''$.
Furthermore, our choice of $r$ implies that the depth of the recursion 
defined by equation~(\ref{eq:cost'}) is on the order of~$\log^\star(N)$, 
so unfolding the recursion gives the theorem.
\end{proof}

\section{Hard problems related to element distinctness}\label{sechardproblems}

In~this section, we consider some related problems for which quantum 
computers cannot improve upon classical (probabilistic) complexity.

\begin{description}
\item[\textbf{Parity-collision problem}]\mbox{}\\
Given function $f:X\rightarrow Z$, find the parity of the cardinality
of the set 
$C_{f}=\{(x,y)\in X \times X \ : \ x<y\mbox{ and } f(x)=f(y)\}$.
\end{description}

\begin{description}
\item[\textbf{No-collision problem}]\mbox{}\\
Given function $f:X\rightarrow Z$, find an element $x\in X$  
that is not involved in a collision (i.e.~$f^{-1}(f(x))=\{x\}$).
\end{description}

\begin{description}
\item[\textbf{No-range problem}]\mbox{}\\
Given function $f:X\rightarrow Z$, find $z\in Z$ such that
$z\not\in f(X)$.
\end{description}

We~assume that $X=Z=[N]$, 
and show that these problems are hard even for
the function-evaluation model.
\begin{theorem}\label{theorem:hard}
The evaluation-complexities of the parity-collision problem,
the no-collision problem and the no-range problem are
lower bounded by $\Omega(N)$.
\end{theorem}

Note that the hardness of the parity-collision problem implies
the hardness of exactly {\em counting} the number of collisions.
Our proofs use the nice lower bound method developed 
by Ambainis~\cite{ambainis:lowerbounds}.
Let us state here exactly the result that we require
(the result is stated in~\cite{ambainis:lowerbounds} 
for inputs which are Boolean vectors, but holds in our case as well).
\begin{theorem}[\cite{ambainis:lowerbounds}]\label{theorem:ambainis}
Let ${\cal F}=\{f:[N]\rightarrow[N]\}$ be the set of all possible
input-functions, and $\Phi:{\cal F}\rightarrow Z$ be a function
(which we want to compute).
Let $A,B$ be two subsets of ${\cal F}$ such 
that $\Phi(f)\neq \Phi(g)$ if $f\in A$ and $g\in B$, 
and $R\subseteq A\times B$ be
a relation such that
\begin{enumerate}
\item For every $f\in A$, there exist at least $m$ different $g\in B$ 
such that $(f,g)\in R$.
\item For every $g\in B$, there exist at least $m'$ different $f\in A$ 
such that $(f,g)\in R$.
\item For every $f\in A$ and $x\in [N]$, there exist at most $l$ 
different $g\in B$ such that $(f,g)\in R$ and $f(x)\neq g(x)$.
\item For every $g\in B$ and $x\in [N]$, there exist at most $l'$ 
different $f\in A$ such that $(f,g)\in R$ and $f(x)\neq g(x)$.
\end{enumerate}
Then any quantum algorithm computing $\Phi$ with probability at least 
$2/3$ requires $\Omega(\sqrt{\tfrac{mm'}{ll'}})$ evaluation-queries.
\end{theorem}

We now give our proof of Theorem~\ref{theorem:hard}.

\medskip

\begin{proof}
To apply Theorem~\ref{theorem:ambainis}, we will 
describe a relation $R$ for each of our problems.
For functions $f:[N]\rightarrow [N]$
and $g:[N]\rightarrow [N]$, we denote by $d(f,g)$ the cardinality of
the set $\{x\in [N] \mid f(x)\neq g(x)\}$. 
For each problem $R$ will be defined by 
$$
R=\{(f,g)\in A\times B \mid d(f,g)=1\},
$$
for some appropriate sets $A$ and $B$.
\begin{description}
\item[Parity-collision problem] Here we suppose that $4$ divides $N$.
Let $A$ be the set of functions $f:[N]\rightarrow [N]$
such that $\abs{C_{f}}=N/4$ and 
$\abs{f^{-1}(z)}\leq 2$ for all $z\in [N]$.
Let $B$ be the set of functions $g:[N]\rightarrow [N]$
such that $\abs{C_{g}}=N/4+1$ and 
$\abs{g^{-1}(z)}\leq 2$ for all $z\in [N]$.
Then a simple computation gives that the relation $R$ satisfies
$m=\Theta(N^{2})$, $m'=\Theta(N^{2})$, $l=\Theta(N)$,
and $l'=\Theta(N)$.
\item[No-collision problem] Now we suppose that $N$ is odd.
Let $A=B$ be the set of functions $f:[N]\rightarrow [N]$
such that $\abs{C_{f}}=(N-1)/2$, and 
$\abs{f^{-1}(z)}\leq 2$, for all $z\in [N]$.
Then $R$ satisfies that
$m=m'=\Theta(N)$ and $l=l'=\Theta(1)$. 
\item[No-range problem]
Let $A=B$ be the set of functions $f:[N]\rightarrow [N]$
such that $C_{f}=\{(1,2)\}$.
Then a similar computation gives
$m=m'=\Theta(N)$ and $l=l'=\Theta(1)$. 
\end{description}
Note that the no-collision problem and the no-range problem are not 
functions in general (several outputs may be valid for one input),
but that they {\em are} functions on the sets $A$ and $B$ chosen above
(there is a unique correct output for each input).
Thus, Theorem~\ref{theorem:ambainis} implies a lower 
bound of $\Omega(N)$ for the evaluation-complexity
of each of our three problems.
\end{proof}

\section{Finding a triangle in a graph}\label{sectrianglefinding}

Finally we consider a related search problem, 
which is to find a triangle in a graph, provided one exists.  
Consider an undirected graph $G=(V,E)$ on $|V|=n$ nodes
with $m=|E|$ edges.
There are $N=\binom{n}{2}$ edge slots
in $E$, which we can query
in a black~box fashion (see also~\cite[Section~7]{bcwz:qerror}).
The triangle-finding problem is:

\begin{description}
\item[\textbf{Triangle-finding problem}]\mbox{}\\
Given undirected graph $G=(V,E)$, find distinct vertices 
$a,b,c\in V$ 
such that $(a,b),$ $(a,c),$ $(b,c)\in E$.
\end{description}

Since there are $\binom{n}{3}<n^3$ triples $a,b,c$, and we can decide whether
a given triple is a triangle using 3 queries, we can use Grover's 
algorithm to find a triangle in $O(n^{3/2})$ queries.
Below we give an algorithm that works more efficiently for sparse graphs.

\begin{algorithm}{Triangle-finder}
\item \label{triangle:edge}
      Use quantum search to find an edge $(a,b)\in E$ among 
      all $\binom{n}{2}$ potential edges.
\item \label{triangle:node}
      Use quantum search to find a node $c\in V$ such 
      that $a,b,c$ is a triangle.
\item \label{triangle:ampl}
     Apply amplitude amplification on 
     steps \mbox{\ref{triangle:edge}--\ref{triangle:node}}.
\end{algorithm}

Step~\ref{triangle:edge} 
takes $O(\sqrt{n^2/m})$ queries and step~\ref{triangle:node} 
takes $O(\sqrt{n})$ queries.
If there is a triangle in the graph, then the probability 
that step~\ref{triangle:edge} 
finds an edge belonging to this specific triangle is $\Theta(1/m)$.
If step~\ref{triangle:edge} 
indeed finds an edge of a triangle, then with probability
at least 1/2, step~\ref{triangle:node} 
finds a $c$ which completes the triangle.
Thus the success probability of 
steps \mbox{\ref{triangle:edge}--\ref{triangle:node}}
is $\Theta(1/m)$
and the amplitude amplification of step~\ref{triangle:ampl} 
requires $O(\sqrt{m})$ iterations.
The total complexity is hence $O((\sqrt{n^2/m}+\sqrt{n})\sqrt{m})$
which is $O(n+\sqrt{nm})$. 
If $G$ is sparse in the sense that $m=|E|\in o(n^2)$, 
then $o(n^{3/2})$
queries suffice. 
Of course for dense graphs our algorithm
will require $\Theta(n^{3/2})$ queries.

We again obtain lower bounds by a reduction from the $\OR$-problem.
Consider an $\OR$-input $X\in\01^{\binom{n}{2}}$ as a graph on $n$ edges.
Let $G$ be the graph obtained from this by adding 
an $(n+1)$-th node and
connecting this to all other $n$ nodes.
Now $G$ has $|X|+n$ edges, and $\OR(X)=1$ if and only if 
 $G$ contains a triangle.
This gives $\Omega(n^2)$ bounds for exact quantum and bounded-error classical,
and an $\Omega(n)$ bound for bounded-error quantum.
The next theorem follows.
\begin{theorem}
If $\Omega(n)\leq |E|\leq \binom{n}{2}$, 
then the edge-query-complexity 
of triangle-finding is
\begin{itemize}
\item $\Omega(n)\leq Q_2(\Triangle)\leq O(n+\sqrt{nm})$
\item $Q_E(\Triangle) = \Theta(n^2)$
\end{itemize}
where $n=|V|$ and $m=|E|$ for $G=(V,E)$.
\end{theorem}

Note that for graphs with $\Theta(n)$ edges, the bounded-error
quantum bound becomes 
$\Theta(n)$ queries, whereas the classical bound remains $\Theta(n^2)$.
Thus we have a quadratic gap for such very sparse graphs.

\section{Concluding remarks}
\label{secohmydearthemoonisagreencheese}
The main problem left open by this research is to close the
gap between upper and lower bounds for element distinctness.
None of the known methods for proving quantum lower bounds seems to 
be directly applicable to improve the $\Omega(\sqrt{N})$ lower bound,
and we feel that if element distinctness is strictly harder than 
unordered search, proving it will require new ideas.

An interesting direction could be to take into account
simultaneously time complexity and space complexity,
as recently has been done for classical 
algorithms by Yao~\cite{yao:timespace}, Ajtai~\cite{ajtai:linear},
Beame, Saks, Sun, and Vee~\cite{bssv:timespace}, and others.
In particular, Yao shows that the time-space product of any classical
deterministic comparison-based branching program 
solving element distinctness 
satisfies $T\cdot S\geq \Omega(N^{2-\eps(N)})$,
where $\eps(N)=5/\sqrt{\ln N}$.
A~possible extension of this result to quantum computation could be
that the time and space of any quantum bounded-error algorithm
for element distinctness satisfies
$T^2\times S \geq \Omega(N^{2-\eps(N)})$, for some
appropriate function $\eps(N)$. 
If such a bound is valid, then our $N^{3/4}$-algorithm is 
near-optimal when the space complexity is bounded by $O(N^{1/2})$
(for our algorithm, the time complexity is the same as the comparison
complexity up to logarithmic factors).

\newcommand{\etalchar}[1]{$^{#1}$}

\end{document}